\documentclass{article}

\usepackage{arxiv}

\usepackage[utf8]{inputenc} 
\usepackage[T1]{fontenc}    
\usepackage{hyperref}       
\usepackage{url}            
\usepackage{booktabs}       
\usepackage{amsfonts}       
\usepackage{nicefrac}       
\usepackage{microtype}      
\usepackage{lipsum}
\usepackage{graphicx}
\usepackage{multirow}
\usepackage{import}

\usepackage{amsmath}
\usepackage{amssymb}
\usepackage{amsthm}
\usepackage{mathrsfs}

\usepackage{algorithm2e}
\usepackage{tikz}
\usepackage[nameinlink]{cleveref}
\usepackage[nolist,nohyperlinks]{acronym}
\usepackage{placeins}
\usepackage{subcaption}
\usepackage{wrapfig}

\newtheorem{theorem}{Theorem}
\newcommand{\subheading}[1]{\noindent\textbf{#1}}
\newcommand{\cev}[1]{\reflectbox{\ensuremath{\vec{\reflectbox{\ensuremath{#1}}}}}}
\DeclareMathOperator{\Ima}{Im}
\newcommand{\institute}[1]{#1}
\newcommand{\email}[1]{\texttt{#1}}

\usepackage[numbers,compress,sort]{natbib}

\title{Representing Edge Flows on Graphs via Sparse Cell Complexes}

\author{
Josef Hoppe\\
\institute{RWTH Aachen University}\\
\email{hoppe@cs.rwth-aachen.de}\And
Michael T. Schaub\\
\institute{RWTH Aachen University}\\
\email{schaub@cs.rwth-aachen.de}
}

\renewcommand{\headeright}{Preprint}
\renewcommand{\undertitle}{A Preprint}

\date{}

\acrodef{GSP}{Graph Signal Processing}
\acrodef{TSP}{Topological Signal Processing}
\begin{document}
\maketitle
\begin{abstract}
    Obtaining sparse, interpretable representations of observable data is crucial in many machine learning and signal processing tasks.
For data representing flows along the edges of a graph, an intuitively interpretable way to obtain such representations is to lift the graph structure to a simplicial complex:
The eigenvectors of the associated Hodge-Laplacian, respectively the incidence matrices of the corresponding simplicial complex then induce a Hodge decomposition, which can be used to represent the observed data in terms of gradient, curl, and harmonic flows.
In this paper, we generalize this approach to cellular complexes and introduce the flow representation learning problem, i.e., the problem of augmenting the observed graph by a set of cells, such that the eigenvectors of the associated Hodge Laplacian provide a sparse, interpretable representation of the observed edge flows on the graph.
We show that this problem is NP-hard and introduce an efficient approximation algorithm for its solution.
Experiments on real-world and synthetic data demonstrate that our algorithm outperforms state-of-the-art methods with respect to approximation error, while being computationally efficient.

\end{abstract}


\section{Introduction}
In a wide range of applications, we are confronted with data that can be described by flows supported on the edges of a graph \cite{mulder2018network,mendes2023signal}.
Some particularly intuitive and important examples include traffic flows within a street network \cite{gh-transportation-networks}, flows of money between economic agents \cite{iori2008network, borgatti2009social}, or flows of data between routers in a computer network \cite{lee2011study}. 
However, many other scenarios in which some energy, mass, or information flows along the edges of a graph may be abstracted in a similar way \cite{billings2021simplicial}.

As is the case for many other setups in machine learning and signal processing~\cite{bengio2013representation,hamilton2017representation,tovsic2011dictionary}, finding a compact and interpretable approximate representation of the overall pattern of such flows is an important task to assess qualitative features of the observed flow data.
In the context of flows on graphs, the so-called (discrete) Hodge-decomposition~\cite{horak2013spectra,lim2020hodge,schaub2020random,barbarossa2020topological,grady2010discrete,aoki2022urban} has recently gained prominence to process such flow signals, as it can be employed to represent any flow on a graph (or more generally, cellular complex) as a sum of a gradient, curl and harmonic components, which can be intuitively interpreted.
This representation of the data may then be used in a variety of downstream tasks~\cite{schaub2021signal}, such as prediction of flow patterns~\cite{roddenberry2019hodgenet,roddenberry2021principled,smith2022physics}, classification of trajectories~\cite{ghosh2018topological,frantzen2021outlier,pokorny2016topological}, or to smooth or interpolate (partially) observed flow data~\cite{jia2019graph,schaub2018flow}.
Furthermore, deep learning approaches on cell complexes that need or infer cells are currently gaining traction~\cite{bodnar2021weisfeiler,giusti2023cell,battiloro2023latent,hajij2020cell}.

Commonly considered mathematical problem formulations to find a compact representation of data are (variants of) sparse dictionary learning~\cite{tovsic2011dictionary}, in which the aim is to find a sparse linear combination of a set of fundamental atoms to approximate the observed data.
Accordingly, such types of dictionary learning problems have also been considered to learn representations of flows on the edges of a graph~\cite{barbarossa2020topological,barbarossa2020topological2,sardellitti2022topological,sardellitti2021topological,roddenberry2022signal}.
Since the Hodge-decomposition yields an orthogonal decomposition of the flows into non-cyclic (gradient) flows and cyclic flows, these signal components can be approximated via separate dictionaries, and as any gradient flow component can be induced by a potential function supported on the vertices of the graph, the associated problem of fitting the gradient flows can be solved via several standard techniques, e.g., by considering the associated eigenvectors of the graph Laplacian.
To find a corresponding representation of the cyclic flows, in contrast, it has been proposed to lift the observed graph to a simplicial or cellular complex, and then identifying which simplices (or more generally, cells) need to be included in the complex to obtain a good sparse approximation of the circular components of the observed flows~\cite{barbarossa2020topological,barbarossa2020topological2,sardellitti2022topological,sardellitti2021topological}.

Such inferred cell complexes can moreover be useful for a variety of downstream tasks, e.g., data analysis based on as neural networks~\cite{bodnar2021weisfeiler,giusti2023cell}.
In fact, even augmenting graphs by adding randomly selected simplices can lead to significant improvements \cite{burns2023simplicial} for certain learning tasks.
Arguably, having more principled selection methods available could thus be beneficial. For instance, the approach of \cite{battiloro2023latent} could potentially be improved by replacing the selection from a pre-defined set of cells with our approach, reducing its input data demands.
Simplicial and cellular complex based representations have also gained interest in neuroscience recently to desribe the topology of interactions in the brain~\cite{giusti2016two}.

However, previous approaches for the inference of cells from edge flow data limit themselves to simplicial complexes or effectively assume that the set of possible cells to be included is known beforehand.
Triangles are not sufficient to model all cells that occur in real-world networks:
For example, grid-like road networks contain virtually no triangles, necessitating the generalization to cell complexes.
In other applications, triangles may be able to approximate longer $2$-cells, which still results in a unnecessarily complex representation.
In this work, we consider a general version of this problem which might be called flow representation learning problem: given a set of edge-flows on a graph, find a lifting of this graph into a regular cell-complex with a small number of 2-cells, such that the observed (cyclic) edge-flows can be well-approximated by potential functions associated to the 2-cells.
As the solution of this problem naturally leads to the construction of an associated cell complex, we may alternatively think of the problem as inferring an (effective) cell-complex from observed flow patterns.

Our main contributions are as follows: \vspace{-\parskip}\vspace{1pt}
\begin{itemize}
    \item We provide a formal introduction of the flow representation learning problem and its relationships to other problem formulations.
    \item We prove that the general form of flow representation learning we consider here is \texttt{NP} hard.
    \item We provide heuristics to solve this problem and characterize their computational complexity.
    \item We demonstrate that our algorithms outperform current state of the art approaches in this context.
\end{itemize}

\subsection{Related work}

\subheading{Finding cycle bases.}
The cycle space of an undirected graph $\mathcal G$ is the set of all even-degree subgraphs of $\mathcal G$.
Note that the cycle space is orthogonal to the space of gradient flows and (for unweighted graphs) isomorphic to the space of cyclic flows.
A lot of research has been conducted on finding both general and specific cycle bases \cite{syslo1979cycle,horton1987polynomial,kavitha2009cycle}.
Our algorithm uses the central idea that the set of all cycles induced by combining a spanning tree with all non-tree edges is a cycle basis.
However, since we aim for a sparse representation instead of a complete cycle basis, this paper has a different focus.
In this paper, you may therefore think of a cycle basis as a set of simple cycles that covers all edges.

\subheading{Graph Signal Processing and Topological Signal Processing.}
The processing of signals defined on graphs has received large attention over the last decade~\cite{shuman2013emerging,ortega2018graph,dong2020graph}.
The extension of these ideas to topological spaces defined via simplicial or cellular complexes has recently gained attention~\cite{barbarossa2020topological,barbarossa2020topological2,sardellitti2022topological,schaub2021signal,schaub2018flow,grady2010discrete,roddenberry2022signal,battiloro2023latent}, with a particular focus on the processing of flows on graphs~\cite{schaub2018flow}.

The problem we consider here is closely related to a sparse dictionary learning problem~\cite{tovsic2011dictionary} for edge-flows. In contrast to previous formulations~\cite{barbarossa2020topological,barbarossa2020topological2,sardellitti2022topological,battiloro2023latent}, we do not assume that set of cells (the dictionary) is given, which creates a more computationally difficult problem we need to tackle.

\subheading{Compressive Sensing}
Compressive Sensing (CS) \cite{candes2006compressive,candes2006stable} may be interpreted as a variant of sparse dictionary learning that finds a sparse approximation from an underdetermined system of equations.
Although different in both methodology and goals, it is noteworthy that it has been successfully applied in the context of graphs \cite{xu2011compressive}.
A CS application of the graph Laplacian \cite{zhu2012graph} indicates that the lifting to a higher-dimensional Hodge Laplacian could also be used in this context.

\subsection{Outline}
The remainder of this article is structured as follows. 
In~\Cref{sec:background}, we provide a brief recap of notions from algebraic topology,  as well as ideas from graph and topological signal processing.
\Cref{sec:problem} then provides a formal statement of the problem considered, followed by our proposed algorithmic solution (see \Cref{sec:approach}).
Our theoretical hardness results are given in~\Cref{sec:theory}.
We demonstrate the utility of our approach with numerical experiments in~\Cref{sec:experiments}, before providing a brief conclusion.

\section{Background and Preliminaries}
\label{sec:background}

In this section, we recap common concepts from algebraic topology and set up some notation.
In this paper we only consider cell complexes with cells of dimension two or lower, so we will only introduce the required parts of the theory.
However, in general, cell complexes have no such limitation, and our methodology can be adapted to also work on cells of higher dimensions~\cite{hatcher2002algebraic}.

\subheading{Cell Complexes.}
At an intuitive level, cell complexes are extensions of graphs that not only have vertices (0-dimensional cells) and edges (1-dimensional cells), but also (polygonal) faces (2-dimensional cells). 
Such faces can be defined by a closed, non-intersecting path (or \textit{simple cycle}) along the graph, such that the path forms the boundary of the polygonal cell.
Simplicial complexes may be seen as a special case of of cell complexes, in which only triangles are allowed as 2-dimensional cells.
We refer to~\cite{hatcher2002algebraic} for a general introduction to algebraic topology.
Our exposition of the background on cell complexes in the whole section below is adapted from~\cite{roddenberry2022signal}.

Within the scope of this paper, a cell complex (CC) $\mathscr{C}$ consists of a set of so-called cells of different dimensions $k \in \{0,1,2\}$.
For our complex $\mathscr{C}$, we denote the set $k$-cells by $\mathscr{C}_k$.
The $k$-skeleton of $\mathscr{C}$ is the cell complex consisting of all $l$-cells in $\mathscr{C}$ with dimension $l \leq k$.
Akin to graphs, we call the elements of $\mathscr{C}_0$ the nodes and denote them by $v_i$ for $i\in 1,\ldots, |\mathscr{C}_0|$.
Analogously, the elements $e_i$ for $i \in 1,\ldots, |\mathscr{C}_1|$ included in $\mathscr{C}_1$ are called the edges of the cell complex, and we call the elements $\theta_i$ for $i \in 1\ldots, |\mathscr{C}_2|$ included in $\mathscr{C}_2$ the polygons of the complex.

\subheading{Oriented Cells.}
To facilitate computations we assign a reference orientation to each edge and polygon within a cell complex.
We use the notation $\Vec{e}_k = [v_i, v_j]$ to indicate the $k$-th oriented edge from node $v_i$ to node $v_j$, and denote its oppositely oriented counterpart as $\cev{e}_k = [v_j, v_i]$.
The $k$-th oriented $2$-cell, labeled as $\Vec{\theta}_k$, is defined by the ordered sequence $\Vec{e}_{i}, \ldots, \Vec{e}_{j}$ of oriented edges, forming a non-intersecting closed path.
Note that within the sequence $\Vec{\theta}_k$ some edges $\cev e_\ell$ may appear opposite their reference orientation.
Any cyclic permutation of the ordered tuple $\Vec{\theta}_k$ defines the same $2$-cell; a flip of both the orientation and ordering of all the edges defining $\vec{\theta}_k$ corresponds to a change in the orientation of the $2$-cell, i.e., $\cev{\theta}_k = [\cev{e}_{j}, \ldots, \cev{e}_{i}]$.

\subheading{Chains and cochains.} Given a reference orientation for each cell, for each $k$, we can define a finite-dimensional vector space $\mathcal{C}_k$ with coefficients in $\mathbb{R}$ whose basis elements are the oriented $k$-cells. 
An element $\varphi_k \in \mathcal{C}_k$ is called a $k$-\emph{chain} and may be thought of as a formal linear combination of these basis elements.
For instance, a $1$-chain may be written as $\varphi_1 = \sum_i a_i \Vec{e}_i$ for some $a_i\in\mathbb{R}$.
We further impose that an orientation change of the basis elements corresponds to a change in the sign of the coefficient $a_i$.
Hence, flipping the orientation of a basis element, corresponds to multiplying the corresponding coefficient $a_i$ by  $-1$, e.g., $a_1\vec{e}_1 = -a_1\cev{e}_1$.
As for any $k$ the space $\mathcal{C}_k$ is isomorphic to $\mathbb{R}^{|\mathscr{C}_k|}$, we may compactly represent each element $\varphi_k \in \mathcal{C}_k$ by a vector ${\mathbf c=(a_1,...,a_{|\mathscr{C}_k|})^\top}$.
Further, we endow each space $\mathcal C_k$ with the standard $\ell_2$ inner product $\langle \mathbf c_1, \mathbf c_2\rangle  = \mathbf c^\top_1 \mathbf c_2$, and thus give $\mathcal{C}_k$ the structure of a finite-dimensional Hilbert space.

The space of \emph{$k$-cochains} is the dual space of the space of $k$-chains and denoted as $\mathcal{C}^k:=\mathcal{C}_k^*$.
In the finite case, these spaces are isomorphic and so we will not distinguish between those two spaces in the following for simplicity.
(Co-)chains may also be thought of as assigning a scalar value to each cell, representing a signal supported on the cells.
In the following, we concentrate on edge-signals on CCs, which we will think of as flows.
These can be conveniently described by cochains and represented by a vector.

\noindent\textbf{Boundary maps.}
Chains of different dimensions can be related via boundary maps $\partial_k: \mathcal{C}_k\rightarrow \mathcal{C}_{k-1}$, which map a chain to a sum of its boundary components.
In terms of their action on the respective basis elements, these maps are defined as:
$\partial_1(\vec{e}) = \partial([v_i,v_j]) = v_i - v_j$ and $\partial_2(\vec{\theta}) = \partial_2([\vec{e}_{i_1}, \ldots,\vec{e}_{i_m}]) = \sum_{j=1}^m \vec{e}_{i_j}$.
Since all the spaces involved are finite dimensional we can represent these boundary maps via matrices $\mathbf{B}_1$ and $\mathbf{B}_2$, respectively, which act on the corresponding vector representations of the chains.
\Cref{fig:cc-illustration} shows a simple CC and its boundary maps.
The dual co-boundary maps ${\partial_k^\top: \mathcal{C}^{k-1} \rightarrow \mathcal{C}^{k}}$, map cochains of lower to higher-dimensions.
Given the inner-product structure of $\mathcal{C}_k$ defined above, these are simply the adjoint maps to $\partial_k$ and their matrix representation is thus given by $\mathbf{B}_1^\top$ and $\mathbf{B}_2^\top$, respectively.

\subheading{The Hodge Laplacian and the Hodge decomposition}
Given a regular CC $\mathscr C$ with boundary matrices as defined above, we define the $k$-th combinatorial Hodge Laplacian~\cite{lim2020hodge,grady2010discrete,schaub2020random} by:
\begin{equation}
    \mathbf{L}_k = \mathbf{B}_k^\top \mathbf{B}_k + \mathbf{B}_{k+1}\mathbf{B}_{k+1}^\top
\end{equation}
Specifically, the $0$-th Hodge Laplacian operator, is simply the graph Laplacian $\mathbf L_0=\mathbf{B}_1\mathbf{B}_1^\top  $ of the graph corresponding to the $1$-skeleton of the CC (note that $\mathbf{B}_0:=0$ by convention).

Using the fact that the boundary of a boundary is empty, i.e., $\partial_k \circ \partial_{k+1}=0$ and the definition of $L_k$, it can be shown that the space of $k$-cochains on $\mathscr C$ admits a so-called Hodge-decomposition \cite{lim2020hodge,schaub2020random,grady2010discrete}:
\begin{equation}
    \mathcal{C}^k= \Ima (\partial_{k+1}) \oplus \Ima (\partial_{k}^\top) \oplus \ker(L_k).
\end{equation}
In the context of $1$-cochains, i.e., flows, this decomposition is the discrete equivalent of the celebrated Helmholtz decomposition for a continuous vector fields \cite{grady2010discrete}.
Specifically, we can create any gradient signal via a $0$-cochain $\phi \in \mathcal{C}^0$ assigning a potential $\phi_i$ to each node $i$ in the complex, and then applying the co-boundary map $\partial_1^\top$.
Likewise, any curl flow can be created by applying the boundary map $\partial_2$ to a $2$-chain $\eta \in \mathcal{C}^2$ of $2$-cell potentials.

Importantly, it can be shown that each of the above discussed three subspaces is spanned by a set of eigenvectors of the Hodge Laplacian.
Namely, the eigenvectors of the \emph{lower Laplacian} $\mathbf L_k^{low} = \mathbf{B}_k^\top \mathbf{B}_k$ precisely span $\Ima (\mathbf{B}^\top_{k})$ (the gradient space); the eigenvectors of the \emph{upper Laplacian} $\mathbf L_k^{up} = \mathbf{B}_{k+1}\mathbf{B}_{k+1}^\top$ span $\Ima (\mathbf{B}_{k+1})$ (curl space), and the eigenvectors associated to zero eigenvalues span the harmonic subspace.

We denote the projection of any edge flow $\mathbf{f} \in \mathcal{C}^1$ into the gradient, curl, or harmonic subspace of $\mathscr{C}$ by $\text{grad}_\mathscr{C}(\mathbf{f}) = \mathbf{B}_1^T(\mathbf{B}_1^T)^\dagger \mathbf{f}$, $\text{curl}_\mathscr{C}(\mathbf{f})= \mathbf{B}_2(\mathbf{B}_2)^\dagger \mathbf{f}$, or $\text{harm}_\mathscr{C}(\mathbf{f})=(\mathbf{I} -\mathbf{L}_1\mathbf{L}_1^\dagger)\mathbf{f}$ respectively. Here $(\cdot)^\dagger$ denotes the Moore-Penrose Pseudoinverse.

\section{Problem Formulation}
\label{sec:problem}

Consider a given a Graph $\mathcal{G}$ with $N \in \mathbb{N}$ nodes and $E \in \mathbb{N}$ edges, which are each endowed with an (arbitrary but fixed) reference orientation, as encoded in an node-to-edge incidence matrix $\mathbf{B}_1$.
We assume that we can observe $s \in \mathbb{N}$ sampled flow vectors $\mathbf{f}_i$, for $i=1,\dots,s$ defined on the edges.
We assemble these vectors into the matrix $\mathbf{F} = [\mathbf{f}_1,\dots,\mathbf{f}_s] \in \mathbb{R}^{E\times s}$.

Our task is now to find a good approximation of $\mathbf{F}$ in terms of a (sparse) set of gradient and curl flows, respectively.
Leveraging the Hodge-decomposition, this problem can be decomposed into two orthogonal problems.
The problem of finding a suitably sparse set of gradient flows can be formulated as a (sparse) regression problem, that aims to find a suitable set of node potentials $\boldsymbol\phi$ such that $\mathbf{B}_1^\top \boldsymbol \phi$ approximates the observed flows under a suitably chosen norm (or more general cost function).
This type of problem has been considered in the literature in various forms \cite{ortega2018graph}. 
We will thus focus here on the second aspect of the problem, i.e., we aim to find a sparse set of circular flows that approximate the observed flows $\mathbf{F}$.
Without loss of generality we will thus assume in the following that $\mathbf{f}_i$ are gradient free flows (otherwise, we may simply project out the gradient component using $\mathbf B_1$).

This task may be phrased in terms of the following dictionary learning problem:
\begin{equation}
    \min_{\boldsymbol{\xi},\mathbf{B}_2} \sum_{i=1}^s \|\mathbf{f}_i - \mathbf{B}_2 \boldsymbol\xi \|_2^2 \quad\text{s.t.}\quad \|\boldsymbol{\xi}\|_0 < k_1, \|\mathbf{B}_2\|_0 <k_2 \text{ and } \mathbf{B}_2 \in \mathcal{B}_2,
\end{equation}
where $\mathcal{B}_2$ is the set of valid edge-to-cell incidence matrices of cell complexes $\mathscr{C}$ whose $1$-skeleton is equivalent to $\mathcal{G}$, and $k_1,k_2$ are some positive chosen integers.
Note that the above problem may be seen as trying to infer a cellular complex with a sparse set of polygonal cells, such that the orginally observed flows have a small projection into the harmonic space of the cell complex --- in other words, we want to infer a cellular complex, that leads to a good sparse representation of the edge flows.

In the following, we thus adopt a problem in which are concerned with the following loss function 
\begin{equation}
    \text{loss}(\mathscr{C},\mathbf{F}) = \|\text{harm}_{\mathscr{C}}(\mathbf{F})\|_F = \sqrt{\sum_{i=1}^{s}\left|\left|\text{harm}_{\mathscr{C}}(\mathbf f_i)\right|\right|_2^2}, \quad\text{s.t.}\quad \mathscr{C} \text{ has $\mathcal{G}$ as $1$-skeleton}
\end{equation}

There are two variants of the optimization problem we look at. 
First, we investigate a variant with a constraint on the approximation loss:
\begin{equation}\label{eq:problem_variant1}\tag{$\mathscr{P}_1$}
    \min_{\mathscr{C}} \left|\mathscr{C}_2\right| \qquad \text{s.t.}\qquad \text{loss}(\mathscr{C}, \mathbf{F}) < \varepsilon
\end{equation}
Second, we consider a variant with a sparsity constraint on the number of $2$-cells:
\begin{equation}\label{eq:problem_variant2}\tag{$\mathscr{P}_2$}
    \min_{\mathscr{C}} \text{loss}(\mathscr{C},\mathbf{F}) \qquad \text{s.t.}\qquad  \left|\mathscr{C}_2\right| \leq n
\end{equation}

Finally, to assess the computational complexity of the problem, we introduce the decision problem:
Given a graph and edge flows, can it be augmented with $n$ cells s.t. the loss is below a threshold $\varepsilon$?

\begin{equation}\tag{DCS}\label{eq:DCS}
    \text{DCS}(\mathcal G,\mathbf{F},n,\varepsilon) := \exists \mathscr{C} \supseteq \mathcal G: \left|\mathscr{C}_2\right| \leq n, \quad \text{loss}(\mathscr{C},\mathbf{F}) < \varepsilon
    ?
\end{equation}

\section{Algorithmic Approach}
\label{sec:approach}

\begin{figure*}[t]
    \fontsize{8pt}{10pt}\selectfont
    \def\svgwidth{\textwidth}
    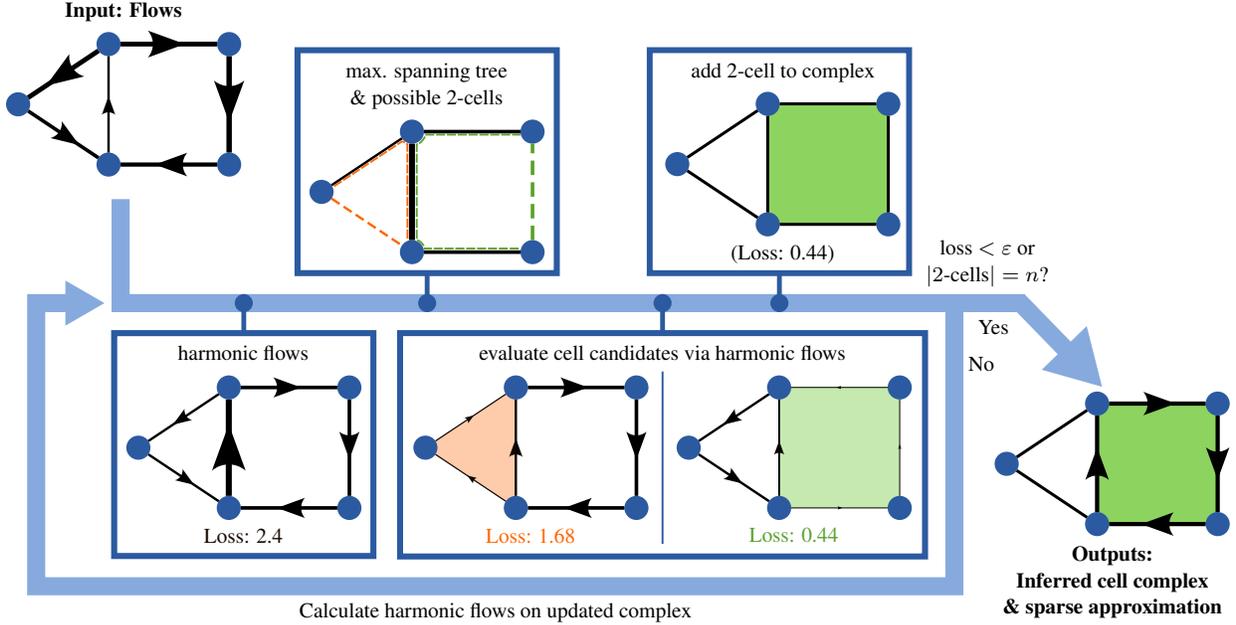
    \caption{Overview of cell complex inference: Our approach starts with a graph and flows supported on its edges. It iteratively adds 2-cells until it fulfills the termination condition. In each iteration, the algorithm projects the flows into the (new) harmonic subspace, selecting cell candidates using a spanning-tree-based heuristic, and adding the cell that minimizes the harmonic flow.}
    \label{fig:1}
    \vspace{-.5cm}
\end{figure*}

We now present a greedy algorithm that approximates a solution for both minimization problems (see~\Cref{fig:1}).
It starts with a cell complex $\mathscr{C}^{(0)}$ equivalent to $\mathcal{G}$ and iteratively adds a new $2$-cell $\theta_i$:

\begin{equation*}
    \mathscr{C}^{(i)} := \mathscr{C}^{(i-1)} \cup \{\theta_i\}, \quad \theta_i := \min_{\theta \in \text{cs}(\mathscr{C}^{(i-1)}, \mathbf{F}, m)} \text{loss}(\mathscr{C}^{(i-1)} \cup \{\theta\}, \mathbf{F})
\end{equation*}

until $i = n$ or $\text{loss}(\mathscr{C}^{(i)}, \mathbf{F}) < \varepsilon$, respectively.
Here, $\text{cs}(\cdot)$ denotes a \textit{candidate search heuristic}, a function that, given a CC and corresponding flows, returns a set of up to $m \in \mathbb{N}$ cell candidates.

\subsection{Candidate search heuristics}
\label{sec:appr-heuristics}

Our algorithm requires a heuristic to select cell candidates because the number of valid cells, i.e., simple cycles, can be in $\Omega(e^{|\mathscr{C}_0|})$ in the worst case (see \cref{sec:apx-heuristics}).
To reduce the number of cells considered, the heuristics we introduce here consider one or a small number of cycle bases instead of all cycles.
Since each cycle basis has a size of $|\mathscr{C}_1| - |\mathscr{C}_0| + 1$, it would be inefficient to construct and evaluate all cycles in the cycle basis.
Instead, we approximate the change in loss via the harmonic flow around a cycle, normalized by the length of the cycle.

Recall that a cycle basis can be constructed from any spanning tree $T$:
Each edge $(u,v) \notin T$ induces a simple cycle by closing the path from $u$ to $v$ through $T$.
Both heuristics efficiently calculate the flow using the spanning tree and select the $m$ cycles with the largest flow as candidates.
Since the selection of spanning trees is so crucial, we introduce two different criteria as discussed below.
\Cref{sec:example-run-appendix} discusses the heuristics in more detail and provides an example of one iteration for each heuristic.

\subheading{Maximum spanning tree.}
The maximum spanning tree heuristic is based on the idea that cycles with large overall flows also have large flows on most edges (when projected into the harmonic subspace).
Since harmonic flows are cyclic flows, the directions tend to be consistent.
However, there may be variations in the signs of the sampled flows $\mathbf{f}_i$.
Therefore, the maximum spanning tree heuristic constructs a spanning tree that is maximal w.r.t.\ the sum of absolute harmonic flows.
See \Cref{alg:max-heuristic} for pseudocode.

\subheading{Similarity spanning trees.}
The maximum spanning tree heuristic does not account for the fact that there might be similar pattern within the different samples.
Given flows $\mathbf{F} \in \mathbb{R}^{E\times s}$, we can represent an edge $e$ using its corresponding row vector $\mathbf{F}_{e,\_}$.
To account for orientation, we insert an edge in both orientations, i.e., both $\mathbf{F}_{e,\_}$ and $-\mathbf{F}_{e,\_}$.
This makes it possible to detect common patterns using $k$-means clustering.
Our similarity spanning trees heuristic exploits this by constructing one spanning tree per cluster center, using the most similar edges.
See \Cref{alg:sim-heuristic} for pseudocode.

\section{Theoretical Considerations}
\label{sec:theory}

\subsection{NP-Hardness of Cell Selection}

\begin{theorem} \label{thm:nph}
    The decision variant of cell selection is NP-hard.
\end{theorem}

We give a quick sketch of the proof here; you can find the complete proof in \cref{sec:np-hard}.

For the proof, we reduce 1-in-3-SAT to DCS.
1-in-3-SAT is a variant of the satisfiability problem in which all clauses have three literals, and exactly one of these literals must be true.

The high-level idea is to represent each clause $c_j$ with a cycle $\gamma_j$, and each variable $x_i$ with two possible cells $\chi_i$ and $\overline{\chi_i}$ containing a long path $\pi_i$ and the clauses that contain $x_i$ and $\overline{x_i}$ respectively.
Through constructed flows, we ensure that every solution with an approximation error below $\varepsilon$ has to

\begin{enumerate}
    \item add either $\chi_i$ or $\overline{\chi_i}$ for every $x_i$, and
    \item contain cells that, combined, cover all clauses exactly once.
\end{enumerate}

This is possible if and only if there is a valid truth value assignment for the 1-in-3-SAT instance.
Consequently, if an algorithm can decide DCS, it can be used to decide 1-in-3-SAT.

\Cref{thm:nph} follows with the NP-Hardness \cite{schaefer1978complexity} of 1-in-3-SAT.

\subsection{Worst-case time complexity of our approach}

The time complexity of one maximum spanning tree candidate search is $O(m \log m)$; for a detailed analysis see \cref{sec:time-complexity-appendix}.

For the similarity spanning trees, having $k$ spanning trees multiplies the time complexity by $k$.
$k$-means also adds an additive component that depends on the number of iterations required for convergence, but is otherwise in $O(nk)$.
Furthermore, $k$-means is efficient in practical applications.

To select a cell from given candidates, we construct $\mathbf{B}_2$ and project the flows into the harmonic subspace.
This computation can be efficiently performed by LSMR \cite{fong2011lsmr} since the matrix is sparse.
However, due to its iterative and numerical nature, a uniform upper bound for its runtime complexity is difficult to obtain.
Instead, we examine the runtime empirically in \Cref{sec:runtime}.

\section{Numerical Experiments}
\label{sec:experiments}

We evaluated our approach on both synthetic and real-world data sets.
To compare our approach to previous work, we adapt the simplicial-complex-based approach from \cite{barbarossa2020topological}.
For this, we exchanged our heuristic based on spanning trees with a heuristic that returns the most significant triangles according to the circular flow around its edges.
Wherever used, this approach is labeled \textit{triangles}.
All code for the evaluation and plotting is available at \url{https://github.com/josefhoppe/edge-flow-cell-complexes}.

When evaluating the sparsity of an approximation, there are conflicting metrics.
Our algorithm optimizes for the definition used in \Cref{sec:problem}, i.e., for a small number of 2-cells, $|\mathscr{C}_2|$.
However, cells with more edges have an inherent advantage over cells with fewer edges simply because the corresponding column in the incidence matrix $\mathbf{B}_2$ has more non-zero entries.
Therefore, we also consider $\|\mathbf{B}_2\|_0$, the number of non-zero entries in $\mathbf{B}_2$, where appropriate.

On synthetic data sets, we also have a ground truth of cells.
We use this information to create a third heuristic (\textit{true\_cells}) that always returns all ground-truth cells as candidates.
Since our approach aims to recover ground-truth cells, we expect true\_cells to outperform it.
If our approach works the way we intend, the difference between it and true\_cells should be relatively small.
For the cell inference problem, we use the ground-truth cells to measure the accuracy of recovering cells.

We construct the cell complexes for the synthetic dataset the following way:

\begin{enumerate}
    \item Draw a two-dimensional point cloud uniformly at random
    \item Construct the Delauney triangulation to get a graph of triangles
    \item Add 2-cells according to parameters by finding cycles of appropriate length
    \item Select edges and nodes that do not belong to any 2-cell uniformly at random and delete them
\end{enumerate}

We construct edge flows $f_i = X_i + \mathbf{B}_2Y_i$ from cell flows $X_i \in \mathcal{C}_2 \sim \mathcal{N}_{\mathscr{C}_2}(0,\mathit{I}\sigma_c)$ and edge noise $Y_i \in \mathcal{C}_1 \sim \mathcal{N}_{\mathscr{C}_2}(0,\mathit{I}\sigma_n)$ sampled i.i.d.\ from multivariate normal distributions with mean $\mu=0$, standard deviation $\sigma_c=1$, and varying standard deviation $\sigma_n \in [0,2]$.

\subsection{Evaluation of cell inference heuristic}
\label{sec:eval-inference-heur}

\begin{figure}[t]
    \centering
    \begin{minipage}[t]{.49\textwidth}
        \centering
        \includegraphics[width=\linewidth]{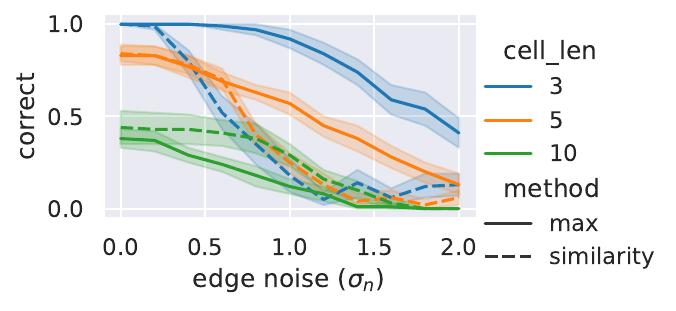}
        \caption{Cell candidates in first iteration. Fraction of correct cells (or combinations thereof); $|\mathscr{C}_2| = 5$; average over 20 runs, 20 flows.}
        \label{fig:heuristic_exp}
    \end{minipage} 
    \hfill
    \begin{minipage}[t]{.49\textwidth}
        \centering
        \includegraphics[width=\linewidth]{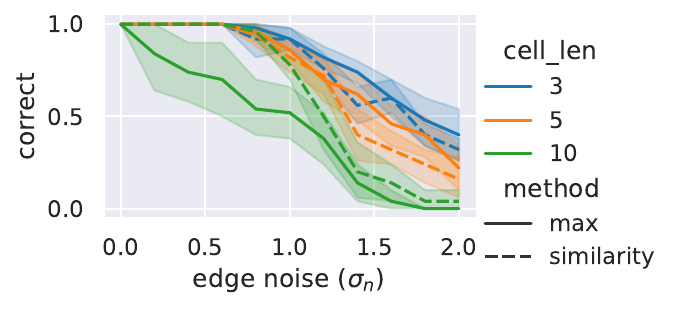}
        \caption{Comparison of inference accuracy of our approach with both heuristics, depending on noise.}
        \label{fig:inference_exp}
    \end{minipage}
    \vspace{-.5cm}
\end{figure}

To evaluate the interpretability of results, we compare them to the ground-truth cells we also used to generate the flows: The cells represent underlying patterns we expect to see in real-world applications.

Before looking at the inference performance of the complete algorithm, we will check that our heuristic works as expected.
\Cref{fig:heuristic_exp} shows that, unsurprisingly, the heuristics work better for shorter cells and if more flows are available.
However, it is not necessary to detect all cells at once as adding one cell results in a new projection into the harmonic space, making it easier to detect further cells.

To evaluate the inference accuracy of the complete algorithm, we determined the percentage of cells detected after $5$ iterations (with five ground-truth $2$-cells to detect).

\Cref{fig:inference_exp} confirms that the overall algorithm works significantly better than the heuristic.
Even for noise with $\sigma_n = 1$, the similarity spanning tree heuristic detects the vast majority of ground-truth cells.
Overall, the experiments on synthetic data indicate that our approach detects underlying patterns, leading to a meaningful and interpretable cell complex.

\subsection{Evaluation of flow approximation quality}

\begin{figure}[t]
    \centering
    \begin{subfigure}[t]{.49\textwidth}
        \centering
        \includegraphics[width=\linewidth]{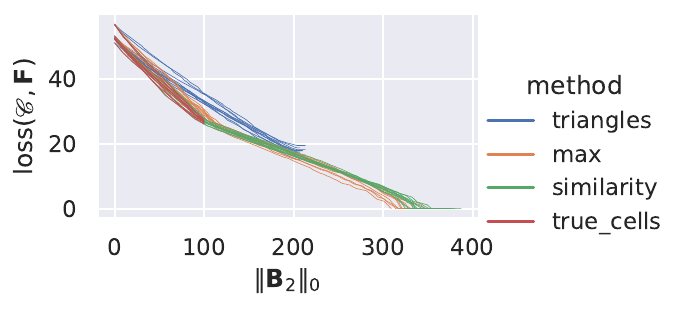}
        \vspace{-.6cm}
        \caption{Approximation error for different sparsity constraints. $\sigma_n=0.75$.}
        \label{fig:approx_error_exp}
    \end{subfigure}
    \hfill
    \begin{subfigure}[t]{.49\textwidth}
        \centering
        \includegraphics[width=\linewidth]{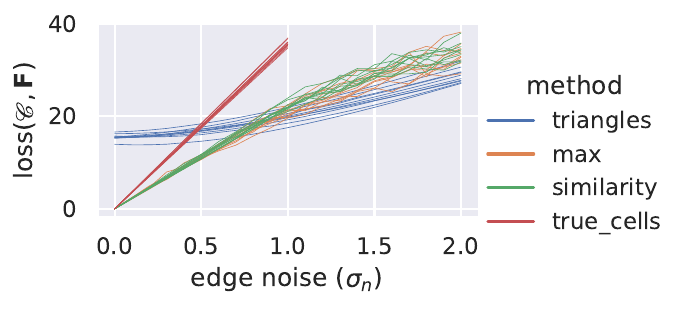}
        \vspace{-.6cm}
        \caption{Approximation error depending $\sigma_n$. Best approximation with $\|\mathbf{B}_2\|_0 \leq 200$; the ground-truth cell complex has a larger error because $\|\mathbf{B}_2^{\text{ground truth}}\|_0 = 100$.}
        \label{fig:approx_error_exp_noise}
    \end{subfigure}
    \vspace{-.2cm}
    \caption{Comparison of our approach, triangles, and ground-truth cells.}
    \vspace{-.5cm}
\end{figure}

As explained before, the triangles heuristic serves as a benchmark representing previous work whereas true\_cells is an idealized version of our approach.

\Cref{fig:approx_error_exp} shows that our approach with the similarity spanning trees heuristic performs close to true\_cells, slightly outperforming the maximum spanning trees, with both significantly outperforming triangles.
Notably, triangles cannot form a complete cycle basis, so only our approach reaches an approximation error of 0.
However, since we are interested in sparse representations, retrieving a complete cycle basis is not our goal.
Instead, we will focus on the behavior for greater sparsity, where the qualitative results depend on the parameter selection.

In general, the longer the cells are, the more significant the difference between the three heuristics becomes.
The approach tends to detect cells with fewer edges than the correct ones in this experiment.
However, smaller cells can be combined to explain the data well for the approximation.
We argue that this is the case with the cells that are found by the algorithm when using the max heuristic:
Compared to true\_cells and similarity, it requires a larger number of cells, but the resulting incidence matrix $\mathbf B_2$ has a similar sparsity.
However, it still outperforms the triangle heuristic, likely because it may take many triangles to approximate a 2-cell.

The amount of noise fundamentally changes the observed behavior, as shown in \Cref{fig:approx_error_exp_noise}, especially when the incidence matrix $\mathbf{B}_2$ is less sparse.
To explain this, we need to look at both the sparsity and dimension of flows.
The vector space of the (harmonic) edge noise has the same dimension as the harmonic space.
Since our approach results in cells with longer boundaries, it reaches the same sparsity with fewer cells than the triangles approach.
With its higher-dimensional approximation, the triangles approach is able to approximate even high-dimensional noise.
If we instead consider the dimension of the approximation $|\mathscr{C}_2|$, our approach outperforms triangles in nearly any configuration with either heuristic (compare \Cref{fig:approx-iter}).

In conclusion, with both sparsity measures, our approach has an advantage for sparse representations.
This observation is consistent with our expectation that the approach can detect the 2-cells of the original cell complex\footnote{Or at least similar cells if the noise makes those more relevant.}.
After detecting the ground-truth cells, the error decreases at a significantly lower rate.
We also expected this change in behavior as the approach now starts to approximate the patterns in the noise, which is bound to be less effective.

\subsection{Experiments on real-world data}

\begin{figure*}[t]
    \centering
    \begin{subfigure}[t]{.48\textwidth}
        \centering
        \includegraphics[width=\linewidth]{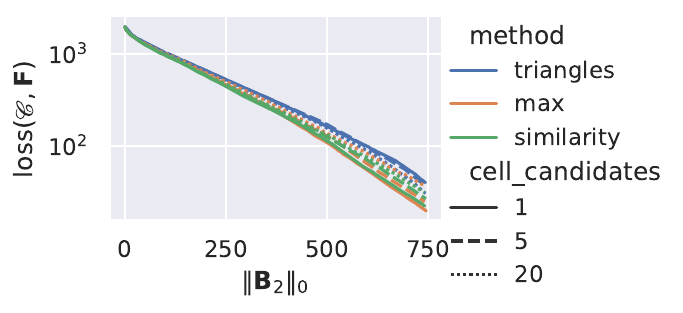}
        \vspace{-.6cm}
        \caption{TransportationNetworks \cite{gh-transportation-networks}: Anaheim.}
        \label{fig:rw_exp_tntp}
    \end{subfigure}
    \hfill
    \begin{subfigure}[t]{.48\textwidth}
        \centering
        \includegraphics[width=\linewidth]{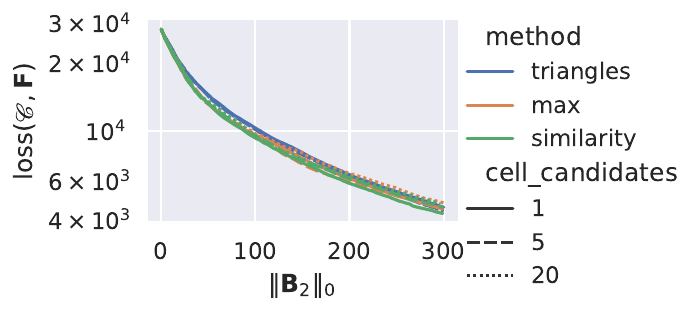}
        \vspace{-.6cm}
        \caption{New York City taxi trips \cite{Benson-2017-spacey,taxi-dataset}.}
        \label{fig:rw_exp_taxi}
    \end{subfigure}
    \vspace{-.2cm}
    \caption{Comparison of our approach and our approach with triangles heuristic. See \Cref{fig:rw-exp-misc} in the appendix for more examples.}
    \label{fig:approx_error_exp_real}
    \vspace{-.5cm}
\end{figure*}

For our evaluation on real-world data, we considered traffic patterns from TransportationNetworks \cite{gh-transportation-networks}, where we extract a single flow per network by calculating the net flow along a link.
For an experiment with multiple flows, we grouped trips of New York City taxis \cite{Benson-2017-spacey,taxi-dataset} and counted the difference in transitions between neighborhoods.

We observe a similar, but less pronounced behavior as in synthetic data.
On the Anaheim network in \Cref{fig:rw_exp_tntp}, we see that our approach consistently outperforms the triangle-based simplicial complex inference.
For the taxi dataset, \Cref{fig:rw_exp_taxi} shows that, like on synthetic data, the triangle based inference performs well as the sparsity decreases.
Note that the apparent effect that more cell candidates lead to a worse performance only exists when measuring sparsity by $\|\mathbf B_2\|_0$ whereas a comparison based on $|\mathscr{C}_2|$ shows a significantly smaller error when considering more candidates in all experiments on real-world data.
Similarly, our approach significantly outperforms a triangle based cell-search heuristic when considering $|\mathscr{C}_2|$.

In addition to its better performance, we believe that general cell-based representations are easier to interpret when analyzing patterns. 
Indeed, the relative success in recovering the correct cells in synthetic data (for real data we don't have a ground truth) and the general good approximation of the flows, may be seen as an indication that cells detected by our approach are more representative of real underlying patterns.
For the taxi example, at 300 entries in $\mathbf B_2$, our heuristic has added $55$ polygonal $2$-cells in the best case, whereas a triangles based inference approach adds one-hundred $2$-cells.
Similar to what we observed on synthetic data, the triangles heuristic can lead to a higher-dimensional approximation that is also inherently better at approximating noise.
Conversely, our approximation is lower-dimensional which may also make it more suitable for de-noising data.

\subsection{Runtime complexity}

\label{sec:runtime}

Finally, we considered the runtime of our algorithm on graphs of different size and generation methods.
Firstly, we randomly generated cell complexes, with four 2-cells each, as described before (\textit{triangulation}).
Secondly, we also generated cell complexes similar to the Watts-Strogatz small-world network construction~\cite{watts1998collective}, but with a fixed probability of $1\%$ for any additional edge and without removing edges on the circle (\textit{smallworld}).
\begin{wrapfigure}{r}{0.5\textwidth}
    \begin{minipage}[t]{.49\textwidth}
        \centering
        \vspace{-.2cm}
        \includegraphics[width=\linewidth]{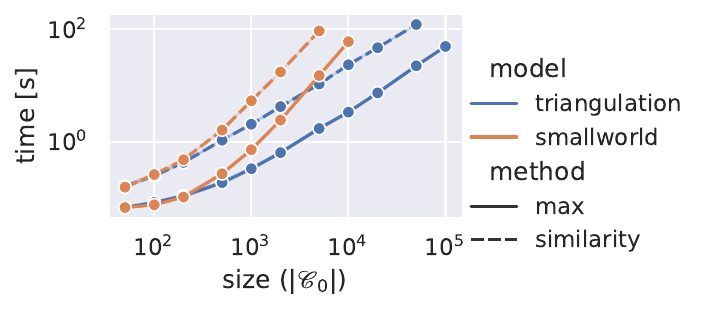}
        \vspace{-.6cm}
        \captionof{figure}{Runtime on graphs of different sizes.}
        \label{fig:time_exp}
    \end{minipage}
\end{wrapfigure}
For the recovery, we generated five synthetic flows and let the algorithm run until it had detected four 2-cells, with five candidates considered in each step.
From our theoretical analysis, we expect the runtime to grow in $O(m\log{}m)$ for the number of edges $m$.
\Cref{fig:time_exp} indicates a slightly superlinear time complexity.
We hypothesize that this stems from the runtime complexity of LSMR, which is hard to assess due to its iterative nature.
In the triangulation graphs, the number of edges is linear in the number of vertices.
In the small-world graphs, the number of edges grows quadratically in the number of vertices, corresponding to faster growth in execution time.
Our algorithm took less than $100s$ for a small-world graph with $10 000$ vertices and a triangulation graph with $100 000$ vertices, respectively.

\section{Conclusion}

We formally introduced the flow representation learning problem and showed that the inherent cell selection problem is NP-hard.
Therefore, we proposed a greedy algorithm to efficiently approximate it.
Our evaluation showed that our approaches surpasses current state of the art on both synthetic and real-world data while being computationally feasible on large graphs.

Apart from further investigation of the inference process and improvements of its accuracy, we see multiple avenues for future research.
A current limitation is that our approach infers cells with a shorter boundary with reasonable accuracy, while cells with a longer boundary have lower inference accuracy.
We may improve this for example by introducing another spanning-tree-based heuristic or de-noising the flows before running it a second time.
On a higher level, the algorithm could be adapted to optimize for sparsity of the boundary map $||\mathbf{B}_2||_0$ instead of the number of two cells $|\mathscr{C}_2|$.

Finally, an analysis of the expressivity of the results on real-world data warrants further investigation.
Given our improvement in approximation over the state of the art, we also expect a better expressivity.
However, since such an analysis does not currently exist, the acutal applicability is hard to assess.
The downstream tasks and improvements to related methods discussed in the introduction could serve as a proxy for this, showing the usefulness of the inferred cells beyond the filtered flow representation.

\section*{Acknowledgements}

Funded by the European Union (ERC, HIGH-HOPeS, 101039827). Views and opinions expressed are however those of the author(s) only and do not necessarily reflect those of the European Union or the European Research Council Executive Agency. Neither the European Union nor the granting authority can be held responsible for them.

\bibliographystyle{unsrtnat}
\bibliography{reference}

\clearpage
\appendix

\section{Cell Complex Illustrations}

\begin{center}
    \captionsetup{type=figure}
    \includegraphics[width=.4\linewidth]{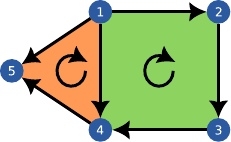}
        \begin{equation*}
            \mathbf{B}_1 = \bordermatrix{ & 1 \rightarrow 2 & 1 \rightarrow 4 & 1 \rightarrow 5 & 2 \rightarrow 3 & 3 \rightarrow 4 & 4 \rightarrow 5 \cr
            1 & -1 & -1 & -1 & 0 & 0 & 0 \cr
            2 & 1 & 0 & 0 & -1 & 0 & 0 \cr
            3 & 0 & 0 & 0 & 1 & -1 & 0\cr
            4 & 0 & 1 & 0 & 0 & 1 & -1\cr
            5 & 0 & 0 & 1 & 0 & 0 & 1} \quad
            \mathbf B_2 = \bordermatrix{ & \text{\includegraphics[height=12pt]{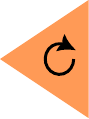}} & \text{\includegraphics[height=12pt]{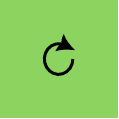}} \cr
      1 \rightarrow 2 & 0 & 1 \cr
      1 \rightarrow 4 & 1 & -1 \cr
      1 \rightarrow 5 & -1 & 0 \cr
      2 \rightarrow 3 & 0 & 1 \cr
      3 \rightarrow 4 & 0 & 1 \cr
      4 \rightarrow 5 & 1 & 0}
        \end{equation*}
\captionof{figure}{A small cell complex and its boundary maps $\mathbf B_1$ and $\mathbf B_2$; columns and rows are annotated with corresponding cells. Flipping the arbitrary orientations of edges and polygons corresponds to multiplying the corresponding rows and columns with $-1$.}\label{fig:cc-illustration}
\end{center}

\section{Heuristics}

In this section, we will discuss heuristics in more detail.
We start with theoretical considerations why heuristics are necessary and what high-level properties they should have.
Then, we provide an example iteration for each heuristic to showcase the most important concepts of each and differences between the heuristics.
Finally, we will discuss advantages and potential pitfalls for each of the heuristics.

\subsection{Theoretical Considerations}
\label{sec:apx-heuristics}

This section discusses the necessity of heuristics and gives some intuition of properties a heuristic should have.
It is supplementary to \cref{sec:appr-heuristics}; for more information on the two heuristics introduced by us, please refer to \cref{sec:example-run-appendix}.

$2$-cells are essentially simple cycles, i.e., cycles without repeating nodes.
Any tuple of nodes that is at least three nodes long can be converted to a cycle.
Since cycles are invariant under shifting and reversing the order of nodes, a cycle of length $l$ can be represented by exactly $2l$ different tuples:
There are $l$ nodes that can be the first node in the tuple.
For each of these, there are two possible orders of the other nodes.

Therefore, on a complete graph, the number of possible two-cells $\mathscr{C}_2^\prime$ is:

\begin{equation}
    |\mathscr{C}_2^\prime| = \sum_{l=3}^{|\mathscr{C}_0|} \frac{1}{2l} \binom{|\mathscr{C}_0|}{l} = \Omega\left(e^{|\mathscr{C}_0|}\right)
\end{equation}

Please not that while this is the worst case, the number of possible $2$-cells is still very large on both randomly generated and real-world graphs with sufficiently many edges; however, this is not trivial to show as the exact number depends heavily on the structure of the given graph.

In general, considering all cells would result in an exponential runtime of our algorithm.
Instead, we introduce heuristics that look at a (relatively speaking) small number of cells.
Accordingly, the heuristic should also not consider all possible cells.
The cells it does consider, however, should include all relevant flows in some meaningful way.

This property is fulfilled by a cycle basis.
A cycle basis spans the subspace of all eulerian subgraphs and is able to model any flow.
While we do not aim to add the complete cycle basis, it can generate a minimal input that considers all flows.
Even though this can theoretically model any harmonic flow, there are cycle bases that are better suited to this task than others.
Since every spanning tree induces a cycle basis, instead of trying to find a suitable cycle basis, we can think of it as finding a good spanning tree.

Furthermore, if the heuristic fully constructs the cycle basis, i.e., enumerates all edges for all elements of the cycle basis, this will result in a realtively high polynomial runtime.
Given that the number of edges may already be quadratic in the number of nodes, this could severely limit the scalability of the approach.
As we show in \cref{sec:time-complexity-appendix}, we can execute one iteration of the whole heuristic in $O(m \log m)$ if we utilize the structure of the spanning tree.

Accordingly, the main difference between the two heuristics we introduce lies in the construction of one or multiple spanning trees.

\subsection{Comparison of the two heuristics in an example run}
\label{sec:example-run-appendix}

\begin{minipage}{\linewidth}
    \captionsetup{type=figure}
    \includegraphics[width=.33\linewidth]{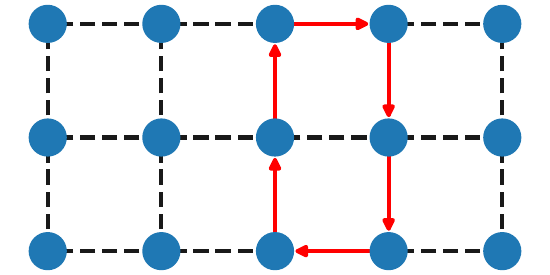}
    \includegraphics[width=.33\linewidth]{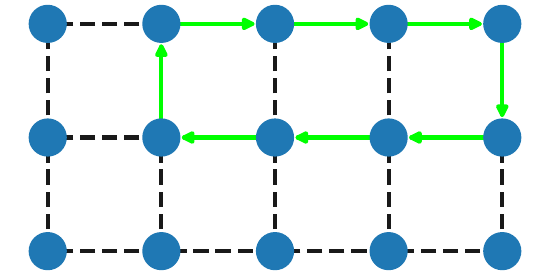}
    \includegraphics[width=.33\linewidth]{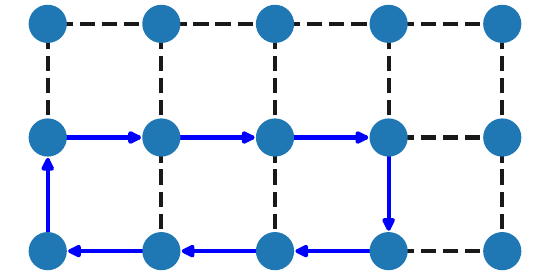}
    \captionof{figure}{The heuristic examples will use a five by three grid graph with three cells. This figure illustrates each cell separately. The cells will be referred to as red, green, and blue; and will be colored accordingly.}
    \label{fig:apx-heur-ex-cc}
\end{minipage}

This section contains a simple example for one iteration of both heuristics presented in this paper.
\Cref{sec:apx-heur-discussion} discusses the importance, advantages, and disadvantages of the heuristics.

\Cref{fig:apx-heur-ex-cc} shows the cell complex we will use for the example; it contains fifteen nodes and three $2$-cells.
Firstly, we generate two flows by assigning a flow to each of the cells and adding noise according to a normal distribution.
Note that the flows in this section are purely for illustrative purposes and do not necessarily reflect the performance on randomly generated synthetic data or real-world data.
The graph structure, $2$-cells, and the flows were deliberatly designed to show a case where the maximum spanning tree heuristic does not result in any ground-truth cell as a candidate while the similarity spanning tree does.
The example flows we use throughout this section are shown in \cref{fig:apx-heur-ex-flows}.

\vspace{2\parskip}
\begin{minipage}{.65\linewidth}
    \captionsetup{type=figure}
    \includegraphics[width=.49\linewidth]{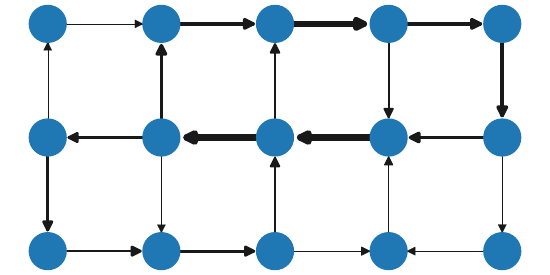}
    \includegraphics[width=.49\linewidth]{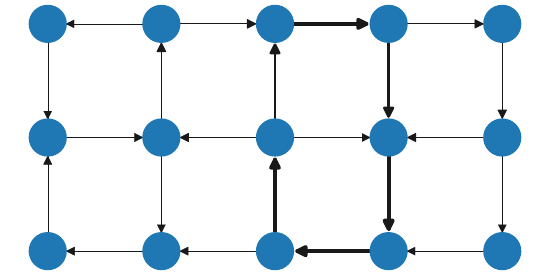}
    \captionof{figure}{The generated flows projected into the harmonic subspace, called flow 1 (left) and flow 2 (right). Note that the blue cell is not visible in either of the flows, making it harder to detect.}
    \label{fig:apx-heur-ex-flows}
\end{minipage}\hfill\begin{minipage}{.32\linewidth}
    \captionsetup{type=figure}
    \includegraphics[width=\linewidth]{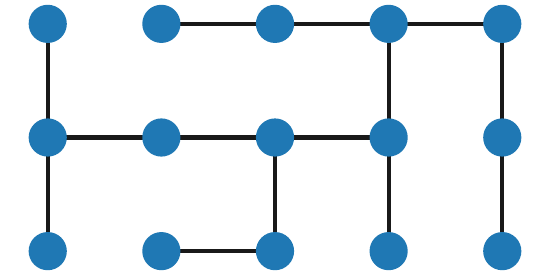}
    \captionof{figure}{The maximum spanning tree as used by the corresponding heuristic.}
    \label{fig:apx-heur-ex-max-st}
\end{minipage}

\subsubsection{Maximum Spanning Tree}

The maximum spanning tree heuristic calculates the total flow value for each edge and constructs a maximum spanning tree, shown in \cref{fig:apx-heur-ex-max-st}.
In this particular example, the maximum spanning tree does not induce any of the three ground-truth cells.
However, it still induces reasonable approximations for both the blue and green cell that each only miss three out of the eight edges.

\subsubsection{Similarity Spanning Tree}

\begin{figure}[t]
    \centering
    \begin{minipage}[t]{\linewidth}
        \raisebox{-0.5\height}{\includegraphics[width=.5\linewidth]{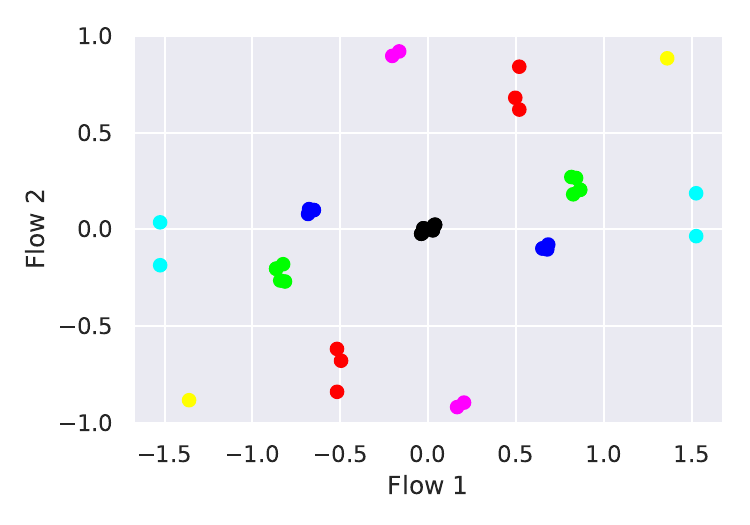}}\hfil
        \raisebox{-0.4\height}{\includegraphics[width=.4\linewidth]{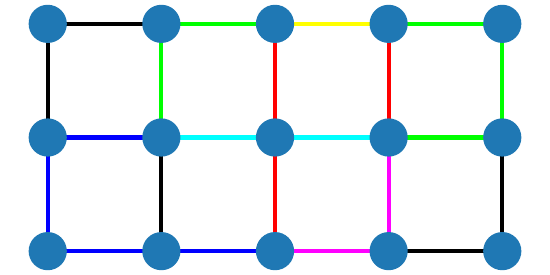}}
    \end{minipage}
    \caption{The edges, represented by their flow values. Because the orientation is arbitrary, each edge appears twice: Once with the original flow values and once with all flow values multiplied by $-1$, representing the possibility of the edge being traversed in the opposite direction. The dots in the scatterplot are colored by combining the colors of all cells they belong to. This coloring is illustrated on the right, where it is applied to the edges in the graph.}
    \label{fig:apx-heur-ex-kmeans}
\end{figure}

\begin{figure}
    \centering
    \includegraphics[width=.32\linewidth]{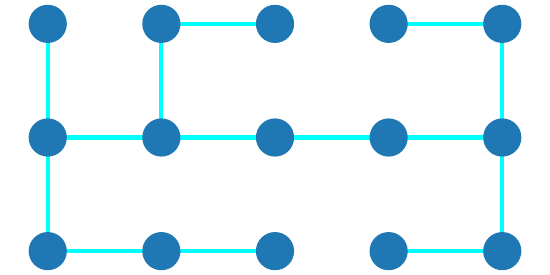}
    \includegraphics[width=.32\linewidth]{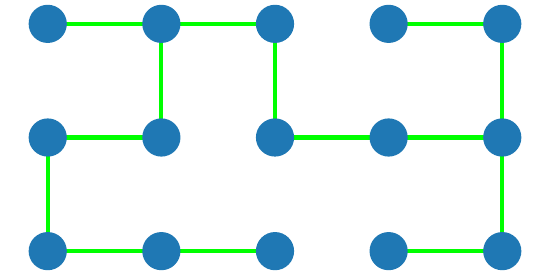}
    \includegraphics[width=.32\linewidth]{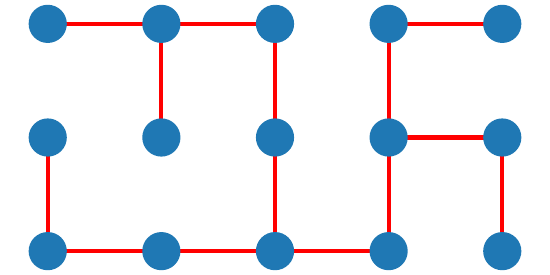}
    \caption{Similarity spanning trees obtained from the cyan (blue + green), green, and red cluster respectively (left to right, colored accordingly).}
    \label{fig:apx-heur-ex-simi}
\end{figure}

The similarity spanning tree heuristic is more complicated.
Firstly, it clusters the edges by their flows.
An optimal clustering can be seen in \cref{fig:apx-heur-ex-kmeans}.
Each of the clusters then induces a similarity spanning tree by adding the edges with the smallest euclidean distance to the cluster center.

Three examples of this are shown in \cref{fig:apx-heur-ex-simi}.
While not every spanning tree induces a ground-truth cell, both the red cell and the green cell are induced by one of these example trees each.
In accordance with a visual examinations of the flows (\cref{fig:apx-heur-ex-flows}), the blue cell is again the hardest to detect.
Since it is also the least significant cell in terms of assigned flow, it is not necessary to detect it as a candidate in the first iterations.
After the first iteration, the most significant cell is added and the flow is re-projected into the harmonic space.
Since this removes interference effects, a future iteration is likely to detect the blue cell.

\subsection{Discussion of heuristics}
\label{sec:apx-heur-discussion}

As the example shows, the selection of appropriate spanning trees is crucial to detecting true cells.
If this does not succeed, the resulting cell complex could still result in a good approximation of the flows.
In the example above, many induced cycles consisted of edges that are part of the same cell.
While this would also like be the case for a random spanning tree on the example graph, an increasing number of edges would make it more crucial to select good edges for a spanning tree.
Should both heuristics fail for a particular application or scale, one could develop an alternative heuristic and easily integrate it to adapt our approach.
With that in mind, we will now discuss potential advantages and pitfalls of the two heuristics we introduced.

\subsubsection{Maximum spanning tree}
As we can see above, the maximum spanning tree heuristic is susceptible to interference effects from adjacent cells.
On the other hand, the heuristic is very fast since it only constructs a single spanning tree (compare \cref{fig:time_exp}) and requires minimal processing of the data.
Furthermore, our experiments in \cref{sec:experiments} show that it provides both good approximations and detects correct cells in many cases (depending on the noise).

\subsubsection{Similarity spanning tree}
The central idea behind the similarity spanning tree heuristic is that edges that are part of the same cell will have similar flow values in all flows.
Since edges can belong to multiple cells, a cell does not correspond to a cluster of edges in general, but will, on average, have a smaller distance to the main cluster than others that do not belong to the cell.
The outperformance of the maximum spanning tree heuristic in \cref{sec:eval-inference-heur} supports this assumption, although it does not always hold.
In \cref{fig:apx-heur-ex-kmeans}, the green cluster is much closer to the blue cluster than any other cluster, and it is even closer to the red cluster than to all points of the cyan cluster.
This results in a spanning tree that does not prefer all edges of the green cell over other edges, as can be seen in \cref{fig:apx-heur-ex-simi}.
However, the cyan cluster results in a spanning tree that induces the green cell, albeit not the blue cell.

In our example, this is the result of very similar flows on the green and blue cells.
Translated to a real-world context, this correlation could indicate a real pattern that may merit further investigation.
In either case, the similarity spanning tree heuristic is able to mitigate this issue by constructing spanning trees from a larger selection of clusters.

Another issue that may arise is the curse of dimensionality if we consider a very large number of flows, rendering the euclidean distance metric less useful.
This is, however, not an inherent property of the heuristic, but more a specific issue exhibited by $k$-means.
Consequently, common mitigation strategies, such as dimensionality reduction using PCR or SVD, may be employed if this becomes an issue for the application of our approach to real-world data.

Other known issues of $k$-means, such as a bad initialization leading to an inaccurate clustering or the difficulty of choosing $k$, are less of a concern:
We are more interested in general trends and even a perfect clustering does not necessarily lead to a perfect result as shown in \cref{fig:apx-heur-ex-kmeans,fig:apx-heur-ex-simi}.

Similarly, the choice of $k$-means is not fixed.
Given its good results and computational performance in a wide variety of fields, we decided to use $k$-means for our implementation.
For the distance metric, we decided to use the euclidean distance because $k$-means also optimizes for it.
Intuitively, many metrics seem appropriate.
However, we specifically decided against angle-based similarity measures as they would give a high similarity to edges that belong to no cluster, but have a similar angle due to noise.
Given the more theoretical and high-level focus of this paper, a thorough comparison of methods for these details would have been out of scope.

\section{Algorithms}
This section provides pseudocode for some of the algorithms mentioned in the main text.
The implementation is available at \url{https://github.com/josefhoppe/cell-flower}.

\begin{algorithm}
    \caption{Maximum spanning tree heuristic. The input consists of the graph $G$, also represented as its set of nodes $V$ and edges $E$; the harmonic flows $F$; and the number of candidates to provide $m$. After calculating the maximum spanning tree, it evaluates the induced cycles and returns the best $m$ cycles as cell candidates. See also \cref{alg:find-tree,alg:build-tree,alg:eval-tree}.}\label{alg:max-heuristic}
    \KwData{$G=(V,E), F, m$}
    $edges \gets [(||F_{\_,e}||_1,e): e \in E]$\;
    $sort(edges)$\;
    $tree\_edges, cycle\_edges \gets \text{find\_spanning\_tree}(edges)$\;
    $p, d, P = \text{spanning\_tree}(V,tree\_edges,F)$\;
    $h \gets Heap()$\;
    $\text{evaluate\_tree}(h,p,d,P,F,cycle\_edges)$\;
    $C \gets \emptyset$\;
    \Repeat{$|C| = m$}{
        $f, p, u, v \gets pop(h)$\;
        $c \gets cycle(p,u,v)$\;
        $C \gets C \cup \{c\}$\;
    }
    \Return{$C$}
\end{algorithm}

\begin{algorithm}
    \caption{Similarity spanning tree heuristic. The input consists of the graph $G$, also represented as its set of nodes $V$ and edges $E$; the harmonic flows $F$; and the number of candidates to provide $m$. After calculating the maximum spanning tree, it evaluates the induced cycles and returns the best $m$ cycles as cell candidates. See also \cref{alg:find-tree,alg:build-tree,alg:eval-tree}.}\label{alg:sim-heuristic}
    \KwData{$G=(V,E), F, k, m$}
    $h \gets Heap()$\;
    $centers \gets k\text{-means}(k, \{\{F_{\_, e} : e \in E\}\})$\;
    \For{$c \in centers$}{
        $edges \gets [(||F_{\_,e} - c||_2,e) : e \in E]$\;
        $sort\_ascending(edges)$\;
        $tree\_edges, cycle\_edges \gets \text{find\_spanning\_tree}(edges)$\;
        $p, d, P \gets \text{spanning\_tree}(V,tree\_edges,F)$\;
        $\text{evaluate\_tree}(h,p,d,P,F,cycle\_edges)$\;
    }
    $C \gets \emptyset$\;
    \Repeat{$|C| = m$}{
        $f, p, u, v \gets pop(h)$\;
        $c \gets cycle(p,u,v)$\;
        $C \gets C \cup \{c\}$\;
    }
    \Return{$C$}
\end{algorithm}

\begin{algorithm}
    \caption{\textit{find\_spanning\_tree.} Given a sorted list edges, find\_spanning\_tree finds a set of edges that form a spanning tree, preferring edges that appear earlier in the list. For example, if the list is sorted descendingly by some edge weight, it returns the edges for the maximum spanning tree.}\label{alg:find-tree}
    \KwData{$edges$}
    $tree\_edges \gets \emptyset$\;
    $cycle\_edges \gets \emptyset$\;
    $uf \gets UnionFind(|V|)$\;
    \For{$(\_, (u,v)) \in edges$}{
        \eIf{$uf(u) \neq uf(v)$}{
            $uf.join(u,v)$\;
            $tree\_edges \gets tree\_edges \cup \{(u,v)\}$\;
        }{
            $cycle\_edges \gets cycle\_edges \cup \{(u,v)\}$
        }
    }
    \Return{$tree\_edges, cycle\_edges$}
\end{algorithm}

\begin{algorithm}
    \caption{\textit{spanning\_tree} constructs a spanning tree from $E$, using the node labeled $0$ as a root. The return values are the parents $p$, the depth in the tree $d$, and the potentials $P$ for all nodes. The potential for a node $n$ is defined as the sum of all flows on the path from the root to $n$.}\label{alg:build-tree}
    \KwData{$V,E,F$}
    $p \gets (-1)_{v\in{}V}$\;
    $p[0] \gets 0$\;
    $d \gets (0)_{v\in{}V}$\;
    $P \gets 0 \in \mathbb{R}^{s\times|V|}$\;
    $q \gets Queue(\{0\})$\;
    \While{$|q|>0$}{
        $v \gets pop(q)$\;
        $N \gets neighbors(V,E,v) \setminus \{p[v]\}$\;
        \For{$u \in N$}{
            $p[u] \gets v$\;
            $d[u] \gets d[v] + 1$\;
            $P_{\_,u} \gets P_{\_,v} + F_{\_,(u,v)}$\;
            $push(q,u)$\;
        }
    }
    \Return{$p,d,P$}
\end{algorithm}

\begin{algorithm}
    \caption{\textit{evaluate\_tree} evaluates all simple cycles induced by a given spanning tree given by the parent list $p$, node depth $d$, and node potentials $P$. The potential for a node $n$ is defined as the sum of all flows on the path from the root to $n$. It calculates the flow of the cycle induced by an edge $(u,v)$ from its flow and the node potentials of $u$ and $v$. The cycle length can be efficiently calculated via the lowest common ancestor \cite{tarjan1979applications}. evaluate\_tree pushes each edge onto the given heap $h$, weighted by its normalized flow.}\label{alg:eval-tree}
    \KwData{$h,p,d,P,F,other\_edges$}
    \For{$(u,v) \in other\_edges$}{
        $f \gets P_{\_,u} - P_{\_,v} + F_{\_,(u,v)}$\;
        $l \gets d[u] + d[v] - 2d[lca(p,u,v)]$\;
        $push(h,(f/l, p, u, v))$
    }
\end{algorithm}

\FloatBarrier

\section{NP-Hardness Proof}
\label{sec:np-hard}

\begin{proof}[Proof of \Cref{thm:nph}]

    To show that the cell selection problem is NP-hard, we reduce 1-in-3-SAT to cell selection.
    1-in-3-SAT is a variant of the satisfiability problem in which all clauses have three literals, and exactly one of these literals must be true.
    1-in-3-SAT is NP-complete \cite{schaefer1978complexity}.
    
    Given an instance $\mathcal{S} = (V,C)$ of 1-in-3-SAT consisting of variables $V = \left\lbrace{}x_1, ..., x_l\right\rbrace$ and clauses $C = \left\{c_1, ..., c_k\right\}$, we now construct an instance of cell selection DCS$(G,F,n,\varepsilon)$.

    In this proof, we use the squared error instead of the $2$-norm because the fact that it is additive simplifies the notation.
    It is possible to apply the square root to $\varepsilon$ and all lower and upper bounds for it with the same qualitative results to show that it also holds for the original definition.
    
    Without limiting generality, we can assume that no clause contains a variable $x_i$ twice, either in positiv or negative form.
    Such a clause $x_i \vee \overline{x_i} \vee x_j$ evaluates to true if and only if $x_j$ is false.
    Therefore, we can remove the clause and the possibility for $x_j$ to be true and continue.
    
    We set $n:=|V|$, i.e., for each variable in $\mathcal{S}$, a cell has to be selected.
    As an intuition, each added cell represents the decision for the value of one $x_i$.
    The literal cell representing $x_i$ ($\overline{x_i}$) is called $\chi_i$ ($\overline{\chi_i}$).
    A clause $c_j \in C$ is represented by a cycle $\gamma_j$ in $G$.
    Analogous to 1-in-3-SAT, the literal cells cover the clause cycles; each $\gamma_j$ has to be covered exactly once in a valid solution.
    We first construct an appropriate base graph $G$.
    Then, we design $2l+1$ flows $F$ on this base graph and select a threshold $\varepsilon$ for the decision problem to ensure that:
    
    \begin{enumerate}
        \item To result in an approximation error below $\varepsilon$, for each $x_i \in V$ either $\chi_i$ or $\overline{\chi_i}$ must be added, and
        \item a solution with approximation error below $\varepsilon$ exists $\Leftrightarrow$ $\mathcal{S}$ is satisfiable.
    \end{enumerate}
    
    We set $p:=2l + 2k + 3$.
    For each variable $x_i$, we create a unique path $\pi_i$ of length $p^7$, including its nodes.
    The first and last nodes on $\pi_i$ are $u_{i,\text{in}}$ and $u_{i,\text{out}}$ respectively.
    We also add a \textit{loop edge} $(u_{i,\text{in}}, u_{i,\text{out}})$ to construct flows later.
    For each clause $c_j = (\alpha \lor \beta \lor \gamma)$ and literal $a \in \{\alpha, \beta, \gamma\}$, we create vertices $v_{j,a,\text{in}}, v_{j,a,\text{out}}$, edges $(v_{j,a,\text{in}}, v_{j,a,\text{out}})$, and paths of length $p$ (inserting new nodes) from $v_{j,\alpha,out}$ to $v_{j,\beta,in}$ etc., thus forming a cycle $\gamma_j$ of length $3p + 3$.
    
    Next, we connect variables to clauses.
    For each $x_i$ and $a \in \left\lbrace x_i, \overline{x_i} \right\rbrace$, let $j_1 < j_2 < ... < j_l$ be the indices of clauses where $a$ occurs.
    We add edges $(u_{i,\text{out}}, v_{j_1,a,\text{in}})$, $(v_{j_k,a,\text{out}}, v_{k_{k+1},a,\text{in}})$ for all $0 < k < l$, and $(v_{i_l,a,\text{out}}, i_{i,\text{in}})$.
    
    \begin{figure}
        \fontsize{8pt}{10pt}\selectfont
        \def\svgwidth{\linewidth}
        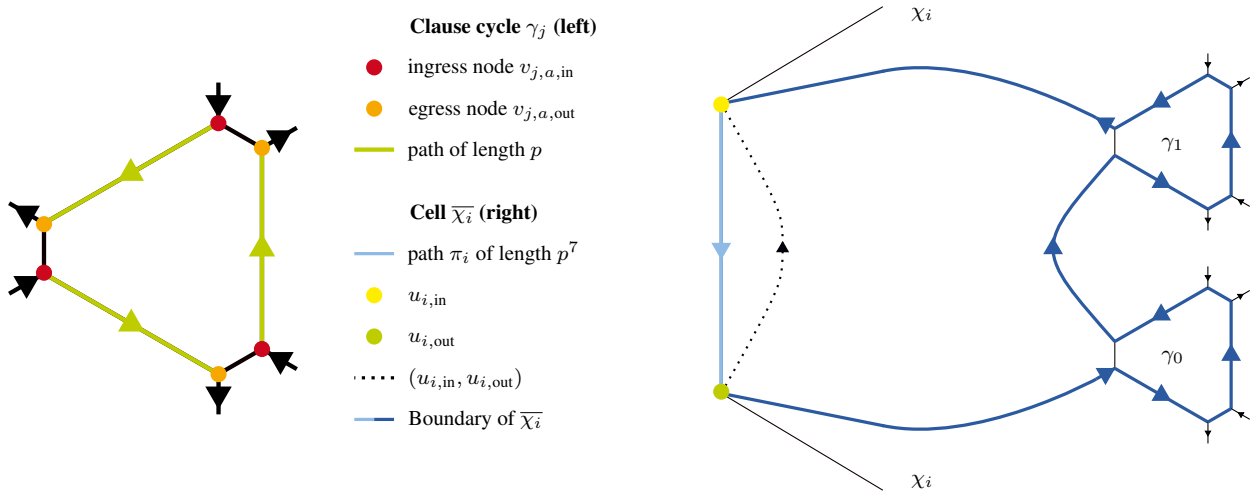
        \caption{Schematic of the Graph for the NP-hardness proof. Arrows indicate the direction of all non-zero flows.}
        \label{fig:np-hardness}
    \end{figure}
    
    With paths $\pi_i$, clause cycles $\gamma_j$, and connecting edges, $G$ is complete.
    A literal $a = x_i$ ($\overline{x_i}$) is represented by a possible cell $\chi_i$ ($\overline{\chi_i}$) with boundary $u_{i,\text{out}} \rightarrow v_{i_1,a,\text{in}} \rightsquigarrow v_{i_1,\beta,\text{out}} \rightarrow v_{i_1,\beta,\text{in}} \rightsquigarrow v_{i_1,\gamma,\text{out}} \rightarrow v_{i_1,\gamma,\text{in}} \rightsquigarrow v_{i_1,a,\text{out}} \rightarrow v_{i_2,a,\text{in}} \rightsquigarrow ... \rightsquigarrow v_{i_p,a,\text{out}} \rightarrow v_{x_i,\text{in}} \rightsquigarrow v_{x_i,\text{out}}$.
    See also \cref{fig:np-hardness} for a visual illustration of a cell representing $\overline{x_i}$ in blue.
    
    In order to analyze the flows, we need to determine an upper bound for the effect the projections into $B_2$ have on the error.
    More specifically: How much can cells that include a $\pi_i$ with flow $0$ affect the overall error, assuming the edges that do not belong to $\pi_i$ have a flow value of at most $\sqrt{p}$.
    There are $2l + 6k + 3kp < p^2$ other edges.
    Therefore, the effect of one cell on one flow is bounded from above by a hypothetical projection with one cell where $p^2$ edges have flow $\sqrt{p}$ and $p^7$ edges have flow $0$.
    In this case, the least squares projection results in a flow of
    
    \begin{align}
            g' &= \frac{\sqrt{p}*p^2+0*p^7}{p^2 + p^7} = \frac{\sqrt{p}}{p^5 + 1}
    \end{align}
    
    If we were to ignore the cell completely, the approximation error is $\sqrt{p}^2\cdot{}p^2+0^2\cdot{}p^6 = p^3$.
    With the cell, the approximation error is
    
    \begin{alignat}{2}
        e &= \left(\sqrt{p}\left(1 - \frac{1}{(p^5+1)^2}\right)\right)^2p^2 + \frac{\sqrt{p}^2}{(p^5+1)^2}p^7
        = \frac{p^{13} + p^8}{(p^5+1)^2}
        = \frac{p^8}{p^5 + 1}
    \end{alignat}
    
    As a result, the reduction in error is
    
    \begin{alignat}{2}
        e' &= p^3 - \frac{p^8}{p^5 + 1}
           = \frac{p^8 + p^3 - p^{8}}{p^5+1}
           = \frac{p^3}{p^5+1}
           < \frac{1}{p^2}
    \end{alignat}
    
    This upper bound for the reduction in error can be summed up over all $n=l<p$ cells and all $2l+1<p$ flows for a total of $\frac{n\cdot{}(2l+1)}{p^2} < 1$.
    In other words, if $z'$ is the error assuming all cells are assigned the flow of their $\pi_i$, the correct projection error $z$ can be bounded by $z'-1 < z \leq z'$.
    
    We will now construct the aforementioned $2|V| + 1$ flows.
    The first $2|V|$ flows ensure that if a solution to $DSC$ exists, its cells are either $\chi_i$ or $\overline{\chi_i}$ for each $x_i \in V$, representing a valid assignment of truth values to variables in $\mathcal{S}$.
    The last flow emulates the evaluation of $\mathcal{S}$ for the given truth value assignment, i.e., the approximation error for this flow is below a certain threshold if and only if the selected cells correspond to a truth value assignment s.t. $\mathcal{S}$ evaluates to true.
    
    To ensure only cells representing literals can be selected with an error smaller than $\varepsilon$, we construct flows $f_{i,1}, f_{i,0} \in C_1$ for each cell $\chi_i$ and $\overline{\chi_i}$.
    It has a positive flow value $\sqrt{p}$) on the boundary of $\chi_i$ ($\overline{\chi_i}$) and a flow of $0$ on all other edges.
    
    Since cell boundaries are cycles, a boundary can either contain all or no edges belonging to each $\pi_i$.
    We observe that for each $\pi_i$, a linear combination of cells that includes $\pi_i$ but not $\pi_j, j\neq{}i$ has to exist:
    If no such linear combination exists, there has to be at least one $i$ where the best approximation of $f_{i,0}$ includes at most $n-1<p$ other $\pi_j, j \neq i$.
    On at least one $\pi_j$, the flow value has to deviate by at least $\frac{\sqrt{p}}{p}$, resulting in an error of at least $\left(\frac{\sqrt{p}}{p}\right)^2 p^7 = p^6 > \varepsilon$.
    If necessary, we can change the basis for the vector space to the linear combinations resulting in each $\pi_i$ belonging to a different cell; therefore, we will assume this from now on.
    
    We can now analyze the cell that includes $\pi_i$ regarding its approximation error on $f_{i,0}, f_{i,1}$.
    If the cell is $\chi_i$ ($\overline{\chi_i}$), it has an approximation error of $||f_{i,0}-f_{i,1}||^2_2$.
    Only $\pi_i$ is shared and the approximation will assign all edges in the boundary a value of $\sqrt{p}$.
    All other edges deviate either on $f_{i,0}$ or on $f_{i,1}$ because the value on $\pi_i$ fixes the flow to $\sqrt{p}$.
    Note that theoretically, this error could be achieved by a cell only covering $\pi_i$ and none of the other edges.
    However, if the cell boundary contains any edge that has a flow value of $0$ in both $f_{i,0}$ and $f_{i,1}$, this will result in an additional approximation error of $\sqrt{p}^2=p$.
    Furthermore, since no variable can occur twice in the same clause, the only vertices shared by $\chi_i$ and $\overline{\chi_i}$ are on $\pi_i$; i.e., the only cells that close $\pi_i$ and don't include an edge that is not covered by $f_{i,0}$ or $f_{i,1}$ are $\chi_i$ and $\overline{\chi_i}$.
    
    We set 
    
    \begin{equation}
    \varepsilon := p - 2 + \sum_{i=1}^{l} ||f_{i,0}-f_{i,1}||^2_2
    \end{equation}
    
    and observe that the error for all $f_{i,0}, f_{i,1}$ is at most $\sum_{i=1}^{l} ||f_{i,0}-f_{i,1}||^2_2 < \varepsilon$ if cells are selected as designed and at least $p - 1 + \sum_{i=1}^{l} ||f_{i,0}-f_{i,1}||^2_2 > \varepsilon$ if they are not (already accounting for the projection of other cells).
    
    Finally, we construct the flow $f_\text{d}$ that mimics the evaluation of $\mathcal{S}$ for a given truth value assignment.
    For this, we set the flow for all variable paths, loop edges, and all clause cycles to $1$.
    Since the variable paths and loop edges form a cycle, grad$(f_{\text{d}}) = 0$.
    
    If $\mathcal{S}$ is satisfiable, i.e., a valid truth value assignment exists, we can use it to construct a solution to DCS.
    We select cells according to the truth values of all variables.
    Since these truth values cover each clause exactly once, the same is true for cells and clause cycles.
    The flow for each cell is $1$ and we can calculate an upper bound for the approximation error:
    For each literal $a \in \left\{x_i, \overline{x_i}\right\}$ that is evaluated to \textit{true}, the corresponding cell has an error of $1$ on
    
    \begin{enumerate}
        \item edges that end in $v_{j,a,in}$ for all clauses $c_j \in C$,
        \item edges $(v_{j,a,out}, v_{j,a,in})$ for all clauses $c_j \in C$, and
        \item one edge per cell that connects the last clause to $v_{x_i,in}$.
    \end{enumerate}
    
    The total number of these edges is $l + 2k$.
    In combination with $l$ the loop edges $(u_{i,\text{out}}, u_{i,\text{in}})$, this results in an upper bound of $2l + 2k$ for $f_\text{d}$ and $ 2l + 2k + \sum_{i=1}^{l} ||f_{i,0}-f_{i,1}||^2_2 < p - 2 + \sum_{i=1}^{l} ||f_{i,0}-f_{i,1}||^2_2 = \varepsilon$ for the error if the 1-in-3-SAT instance is satisfiable.
    
    If $\mathcal{S}$ is not satisfiable, every solution has at least one clause that is not covered or covered at least twice.
    In both cases, the approximation is off by at least $1$ on every edge of the cell, resulting in an error of $3p + 3$ on $f_\text{d}$.
    Overall, this results in a lower bound for the error of $3p + 3 - 1 + \sum_{i=1}^{l} ||f_{i,0}-f_{i,1}||^2_2 > p - 2 + \sum_{i=1}^{l} ||f_{i,0}-f_{i,1}||^2_2 = \varepsilon$ (already accounting for the projection of other cells).
    Therefore, DCS$(G,F,n,\varepsilon)$ is equivalent to $\mathcal{S}$.
    
    By using this reduction and the NP-Completeness of 1-in-3-SAT, we have shown that the decision variant of the cell inference problem is NP-Hard.
\end{proof}

\FloatBarrier
\clearpage

\section{Time Complexity for Heuristics}
\label{sec:time-complexity-appendix}

The time complexity of one maximum spanning tree candidate search is $O(m \log m)$:
Using the Union-Find data structure, we can construct a maximum spanning tree in $O(m \log m)$ by first sorting the edges by total flow and then subsequently adding edges iff their nodes are not already connected (using UnionFind \cite{tarjan1979applications} in $O(\alpha(n))$ per edge\footnote{Where $\alpha$ is the inverse of the Ackermann function}).
For each node, we calculate its \textit{flow potential} as the sum of all edge flows on the path between it and the root (considering the direction).
To get the total flow along a cycle induced by a new edge $(u,v)$, we add the flows for the edge to the difference between the potentials at $u$ and $v$.
The length of the cycle induced by $(u,v)$ is $d(u) + d(v) - 2*d(\text{lca}(u,v)) + 1$, where $d(u)$ is the depth of node $u$ in the tree and $\text{lca}$ denotes the lowest common ancestor of two nodes.
The length of all induced cycles can be obtained in $O(m \alpha(m))$ by using Tarjan's off-line lowest common ancestors algorithm \cite{tarjan1979applications}.

\section{Additional Numerical Experiments}
\begin{figure}[ht]
    \centering
    \begin{subfigure}[t]{.49\textwidth}
        \centering
        \includegraphics[width=\linewidth]{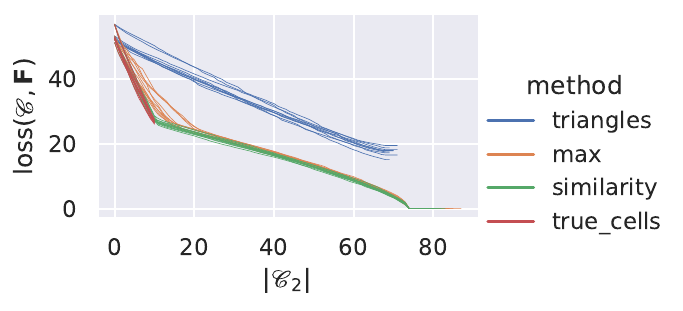}
        \caption{Approximation error for different sparsity constraints. Noise with $\sigma=0.75$.}
        \label{fig:approx_error_exp_iter}
    \end{subfigure}
    \hfill
    \begin{subfigure}[t]{.49\textwidth}
        \centering
        \includegraphics[width=\linewidth]{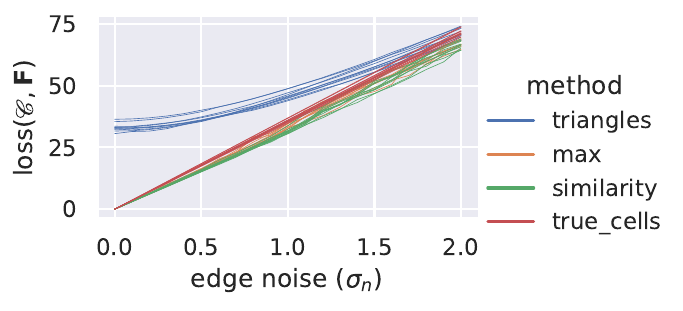}
        \caption{Approximation error for noise with different $\sigma$. Best approximation with $|\mathscr{C}_2| \leq 20$; the ground-truth cell complex has a larger error because $|\mathscr{C}_2| = 10$.}
        \label{fig:approx_error_exp_noise_iter}
    \end{subfigure}
    \caption{Comparison of our approach, triangles, and ground-truth cells. Sparsity measured by $|\mathscr{C}_2|$.}
    \label{fig:approx-iter}
\end{figure}

\begin{figure}[ht]
    \centering
    \begin{subfigure}[t]{.49\textwidth}
        \centering
        \includegraphics[width=\linewidth]{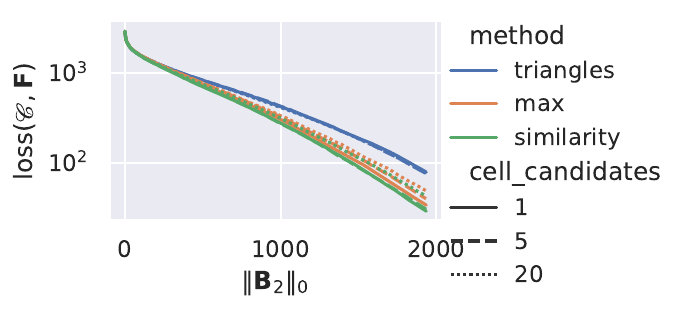}
        \caption{Barcelona.}
    \end{subfigure}
    \hfill
    \begin{subfigure}[t]{.49\textwidth}
        \centering
        \includegraphics[width=\linewidth]{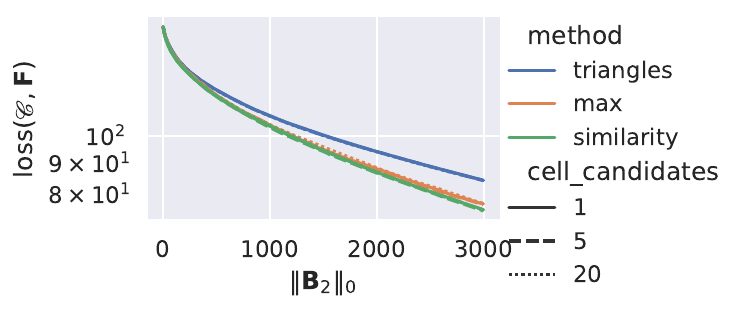}
        \caption{Berlin-Center.}
    \end{subfigure}
    \begin{subfigure}[t]{.49\textwidth}
        \centering
        \includegraphics[width=\linewidth]{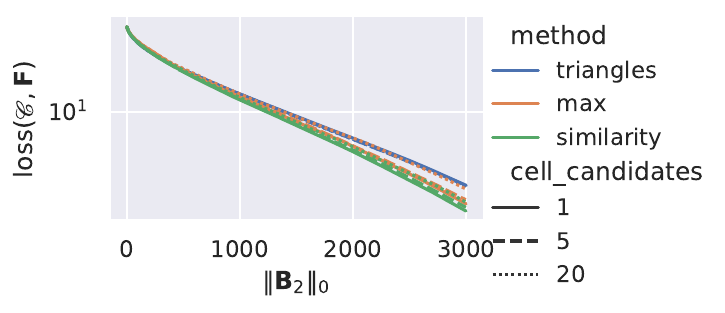}
        \caption{Berlin-Mitte-Prenzlauerberg-Friedrichshain-Center.}
    \end{subfigure}
    \hfill
    \begin{subfigure}[t]{.49\textwidth}
        \centering
        \includegraphics[width=\linewidth]{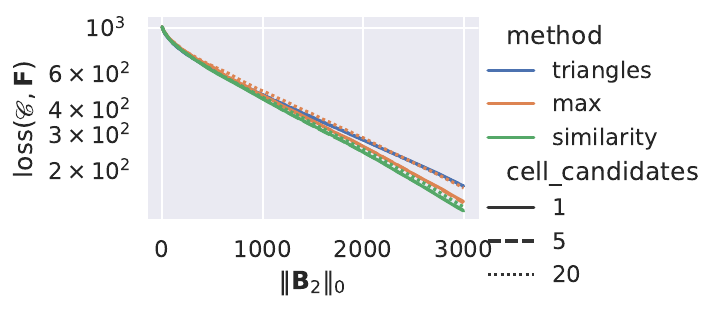}
        \caption{Winnipeg.}
    \end{subfigure}
    \caption{Experiments on TransportationNetworks \cite{gh-transportation-networks} datasets.}
    \label{fig:rw-exp-misc}
\end{figure}

\end{document}